\documentclass[preprint,aps]{revtex4}

\usepackage{graphicx}

\begin{document}

\title{Common Electronic Origin of Superconductivity in (Li,Fe)OHFeSe Bulk Superconductor and Single-Layer FeSe/SrTiO$_3$  Films}

\author{Lin Zhao$^{1,\sharp}$, Aiji Liang$^{1,\sharp}$, Dongna Yuan$^{1,\sharp}$, Yong Hu$^{1,\sharp}$, Defa Liu$^{1}$, Jianwei Huang$^{1}$, Shaolong He$^{1}$, Bing Shen$^{1}$, Yu Xu$^{1}$, Xu Liu$^{1}$, Li Yu$^{1}$, Guodong  Liu$^{1}$, Huaxue Zhou$^{1}$, Yulong Huang$^{1}$, Xiaoli Dong$^{1,*}$, Fang Zhou$^{1}$, Zhongxian Zhao$^{1,2}$, Chuangtian Chen$^{3}$, Zuyan Xu$^{3}$ and X. J. Zhou$^{1,2,*}$ }

\affiliation{
\\$^{1}$National Lab for Superconductivity, Beijing National Laboratory for Condensed Matter Physics, Institute of Physics,
Chinese Academy of Sciences, Beijing 100190, China
\\$^{2}$Collaborative Innovation Center of Quantum Matter, Beijing 100871, China
\\$^{3}$Technical Institute of Physics and Chemistry, Chinese Academy of Sciences, Beijing 100190, China
\\$^{\sharp}$These authors contributed equally to the present work.
\\$^{*}$Corresponding author: XJZhou@aphy.iphy.ac.cn, dong@aphy.iphy.ac.cn
}
\date{May 23, 2015}
%
%


\maketitle

{\bf  The mechanism of high temperature superconductivity in the iron-based superconductors remains an outstanding issue in condensed matter physics\cite{Stewart,FWang,Dagotto}. The electronic structure, in particular the Fermi surface topology, is considered to play an essential role in dictating the superconductivity\cite{Kuroki,Mazin1,FWang1,AVChubukov,VStanev,FWang2,Mazin2}. Recent revelation of distinct electronic structure and possible high temperature superconductivity with a transition temperature T$_c$ above 65 K in the single-layer FeSe films grown on the SrTiO$_3$ substrate provides key information on the roles of Fermi surface topology and interface in inducing or enhancing superconductivity\cite{QKXue,DFLiu,SLHe,SYTan,WHZhangCPL,JJLee,XLiu,JFHe,RPeng1,RPeng2,JFGe,XLReview}. Here we report high resolution angle-resolved photoemission measurement on the electronic structure and superconducting gap of a novel FeSe-based superconductor, (Li$_{0.84}$Fe$_{0.16}$)OHFe$_{0.98}$Se,  with a T$_c$ at 41 K.  We find that this single-phase bulk superconductor shows remarkably similar electronic behaviors to that of the superconducting single-layer FeSe/SrTiO$_3$ film in terms of Fermi surface topology, band structure and nearly isotropic superconducting gap without nodes. These observations provide significant insights in understanding high temperature superconductivity in the single-layer FeSe/SrTiO$_3$ film in particular, and the mechanism of superconductivity in the iron-based superconductors in general.}




The iron-based superconductors\cite{Stewart,FWang,Dagotto} represent the second class of high temperature superconductors after the first discovery of high temperature superconductivity in cuprates\cite{Bednorz}. The electronic structure of the iron-based compounds usually consists of hole-like bands near the Brillouin zone center and electron-like bands near the Brillouin zone corners, leading to proposals that the electron scattering between the hole pockets around the Brillouin zone center and the electron pockets around the Brillouin zone corners is a viable mechanism for electron pairing in the iron-based superconductors\cite{Kuroki,Mazin1,FWang1,AVChubukov,VStanev,FWang2,Mazin2}. With the discovery of the A$_x$Fe$_{2-y}$Se$_2$ (A=K, Cs, Rb, Tl and etc.) superconductors\cite{JGGuo,MHFang,DXReview,Dagotto}, such a Fermi surface nesting picture is challenged because the superconducting phase in A$_x$Fe$_{2-y}$Se$_2$ superconductors does not contain hole pockets near the Brillouin zone center\cite{TQian,DXMouPRL,YZhang,LZhaoPRB,DXReview,Dagotto}. The latest discovery of possible high temperature superconductivity above 65 K in the single-layer FeSe film grown on the SrTiO$_3$ substrate (denoted as FeSe/SrTiO$_3$ film hereafter) puts even stronger constraints on the Fermi surface nesting picture because there is no Fermi surface present near the zone center at all\cite{QKXue,DFLiu,SLHe,SYTan,XLReview}.  While the implications of the Fermi surface topology in the A$_x$Fe$_{2-y}$Se$_2$ superconductors and superconducting single-layer FeSe/SrTiO$_3$ films are significant, the material-specific complications make the conclusion on the role of Fermi surface topology ambiguous. The A$_x$Fe$_{2-y}$Se$_2$ superconductors are well-known to have problems of inhomogeneity, phase separation and small superconducting volume fraction. The exact nature of the superconducting phase remains unclear, and its coexistence with the insulating phase suggests that interface may play an important role in the superconductivity\cite{YJYan,WBao,ZWang,WLi}. There are strong indications that interface in the single-layer FeSe/SrTiO$_3$ films plays an important role in giving rise to high temperature superconductivity\cite{QKXue,DFLiu,SLHe,SYTan,WHZhangCPL,JJLee,XLiu,JFHe,RPeng1,RPeng2,JFGe,XLReview}. With the complications of the interface effect and phase separation problem involved in A$_x$Fe$_{2-y}$Se$_2$ superconductors and superconducting single-layer FeSe/SrTiO$_3$ films, it becomes less straightforward to conclude whether the same effect of Fermi surface topology on superconductivity is operative in all the iron-based superconductors.

In this paper, we report high resolution angle-resolved photoemission (ARPES) measurements on a new FeSe-based superconductor, (Li$_{1-x}$Fe$_x$)OHFe$_{1-y}$Se, with a superconducting critical temperature at 41 K\cite{XFLuPRB,XFLuNM,XLDong,XLDong11111,BLei}. We find that this superconductor contains electron-pocket(s) near the Brillouin zone conners without hole-pockets near the zone center. The superconducting gap around the electron-like Fermi surface near the zone corner is nearly isotropic without nodes, and its temperature dependence follows the usual BCS form.  These behaviors are strikingly similar to that of the superconducting single-layer FeSe/SrTiO$_3$ films. This is the first single-phase bulk superconductor with a relatively high T$_c$$\sim$41 K, free from the complication of phase separation and interface effect, that shows electron-like Fermi pockets only.  Our observations provide significant insights on the superconductivity mechanism in the iron-based bulk superconductors, as well as the origin of high temperature superconductivity in the single-layer FeSe/SrTiO$_3$ films.

Figure 1a shows the crystal structure  of the new FeSe-based superconductor: (Li$_{1-x}$Fe$_x$)OHFe$_{1-y}$Se\cite{XFLuNM,XLDong11111}. It consists of FeSe layers sandwiched in between the (LiFe)OH layers along the {\it c} direction. The in-plane lattice constant {\it a} or {\it b} is 3.78 {\AA} which is close to that (3.76 {\AA}) in the bulk FeSe superconductor (T$_c$$\sim$ 8.5 K)\cite{FCHsu}, but the distance of the two adjacent FeSe layers in (Li$_{1-x}$Fe$_x$)OHFe$_{1-y}$Se (9.32 {\AA})  is much larger than that in bulk FeSe (5.5 {\AA})\cite{FCHsu}. This indicates a weak interaction between the two adjacent FeSe layers in (Li$_{1-x}$Fe$_x$)OHFe$_{1-y}$Se and its enhanced two-dimensional nature of the electronic structure compared with bulk FeSe. In comparison, the single-layer FeSe/SrTiO$_3$  film consists of a strictly two-dimensional Se-Fe-Se layer on the top of the SrTiO$_3$ substrate; the thickness of the FeSe layer is  5.5 $\AA$ (Fig. 1b)\cite{QKXue}. In comparison, the distance between the two adjacent FeSe layers in the A$_x$Fe$_{2-y}$Se$_2$ is 7.02 $\AA$, however, there is an in-plane displacement between the two adjacent FeSe layers (Fig. 1c)\cite{JGGuo}.  The latest success of growing high-quality large-sized (Li$_{1-x}$Fe$_x$)OHFe$_{1-y}$Se single crystals\cite{XLDong11111} makes it possible to carry out ARPES measurements on the new superconductors. The sample we have measured in this paper has a composition of Li$_{0.84}$Fe$_{0.16}$OHFe$_{0.98}$Se (abbreviated as FeSe11111 hereafter) \cite{XLDong11111} with a superconducting transition temperature T$_c$$\sim$41 K (Fig. 3d).

Figure 1d shows the Fermi surface mapping of FeSe11111 measured at 20 K. The spectral weight at each momentum is obtained by integrating the EDC (energy distribution curve) spectral weight within [-10,+10] meV energy window with respect to the Fermi level (E$_F$). The measured Fermi surface contains  a nearly circular  electron-like Fermi surface around M points (Fig. 1d).  There is no indication of Fermi crossing around the $\Gamma$ (0,0) point. Such a Fermi surface topology is very similar to that found in the superconducting single-layer FeSe/SrTiO$_3$ film (Fig. 1e) with a superconducting temperature at $\sim$55 K\cite{DFLiu}. In comparison, there are electron pockets around the Brillouin center in the A$_x$Fe$_{2-y}$Se$_2$ superconductor (Fig. 1f)\cite{DXMouPRL,LZhaoPRB}. The area of the electron pocket near M gives a good measure of the electron doping level for the strongly two-dimensional systems.  Considering that the Fermi surface around M consists of two degenerate Fermi surface sheets, the estimated electron counting in FeSe11111 (Fig. 1d) is $\sim$0.08 electrons/Fe.  This is smaller than 0.10 electrons/Fe in single-layer FeSe film (Fig. 1e) with a T$_c$$\sim$55 K\cite{DFLiu} and 0.12 electrons/Fe in vacuum-annealed single-layer FeSe/SrTiO$_3$ film with a T$_c$$\sim$ 65 K\cite{SLHe}.  We note that the electron counting in both FeSe11111 and superconducting single-layer FeSe/SrTiO$_3$ film is much smaller than 0.18 electrons/Fe in A$_x$Fe$_{2-y}$Se$_2$ superconductor (Fig. 1f)\cite{TQian,DXMouPRL,YZhang,LZhaoPRB} when only considering  the M point electron Fermi surface sheet.

Figure 2 shows the band structure (a-c) and photoemission spectra (l-n) of FeSe11111 measured across three momentum cuts. For comparison, the band structure measured on a superconducting single-layer FeSe/SrTiO$_3$ film\cite{DFLiu} (Fig. 3(e-g)) and on (Tl,Rb)$_x$Fe$_{2-y}$Se$_2$ superconductor\cite{DXMouPRL} (Fig. 3(h-j)) along the same three momentum cuts are also presented. Overall, the band structure of FeSe11111 is very similar to that of the superconducting single-layer FeSe/SrTiO$_3$ film with slight quantitative difference in the band position.  Near the $\Gamma$ point, two hole-like bands can be resolved for FeSe11111 (Fig. 2c) that are well below the Fermi level, as seen more clearly in the corresponding second-derivative image (Fig. 2d) and the photoemission spectra (Fig. 2n). We note that there is a very weak signal near the Fermi level at $\Gamma$ point which may be due to the leakage from a band above the Fermi level.  Similar multiple hole-like bands have also been observed near $\Gamma$ in the superconducting single-layer FeSe/SrTiO$_3$ films\cite{SLHe} and in K$_x$Fe$_{2-y}$Se$_2$ superconductor\cite{LZhaoPRB}, with slight variation in the band top position. In comparison, the electronic structure of (Tl,Rb)$_x$Fe$_{2-y}$Se$_2$ superconductor shows two electron-like bands crossing the Fermi level (Fig. 2j)\cite{DXMouPRL}.  Around M2 and M3 points, we observed in  FeSe11111 parabolic electron-like bands crossing the Fermi level (Figs. 2a and 2b), forming the electron pocket around M. These are very similar to those observed in the superconducting single-layer FeSe/SrTiO$_3$ film  (Figs. 2e and 2f). Careful examination indicates that the Fermi momentum (k$_F$) for the electron pocket of FeSe11111 is $\sim$0.22 $\pi$/a (lattice constant a=3.78 $\AA$) (Figs. 2a and 2b) which is slightly smaller than 0.25 $\pi$/a in the superconducting single-layer FeSe/SrTiO$_2$ film (Figs. 2e and 2f)\cite{DFLiu}, both being significantly smaller than 0.35 $\pi$/a in (Tl,Rb)$_x$Fe$_{2-y}$Se$_2$ superconductor ((Figs. 2h and 2i) \cite{DXMouPRL}.  The bottom of the electron-like band in FeSe11111 is $\sim$50 meV below the Fermi level(Fig. 2b), slightly shallower than the 60 meV in the superconducting single-layer FeSe/SrTiO$_3$  Film (Fig. 2f)\cite{DFLiu} but similar to the $\sim$50 meV in the (Tl,Rb)$_x$Fe$_{2-y}$Se$_2$ superconductor (Fig. 2i)\cite{DXMouPRL}.  The combined information of the Fermi momentum and the band width gives a good measure of the electron effective mass that is 2.9m$_{e}$, 2.7m$_{e}$ and 6.1m$_e$  (m$_{e}$ is free electron mass) for FeSe11111, superconducting single-layer FeSe/SrTiO$_3$ film, and the (Tl,Rb)$_x$Fe$_{2-y}$Se$_2$ superconductor, respectively\cite{DFLiu,DXMouPRL}. It is interesting to note that electron effective mass for the electron-like band near M is similar for FeSe11111 and superconducting single-layer FeSe/SrTiO$_3$ film, but both are much smaller than that in (Tl,Rb)$_x$Fe$_{2-y}$Se$_2$ superconductor\cite{DXMouPRL}.

The clear identification of electron-like Fermi surface sheet near M point makes it possible to investigate the superconducting gap in this new FeSe11111 superconductor. We start by  examining the temperature dependence of the superconducting gap.  Fig. 3b shows the original photoemission image measured at a low temperature of 18 K along the momentum cut (its location is shown in Fig. 3a) crossing the electron-like Fermi surface around M2.  The data is divided by the corresponding Fermi distribution function; the spectral weight suppression near the Fermi level indicates the opening of the superconducting gap (Fig. 3b).  Following the procedure commonly used in the study of high temperature cuprate superconductors\cite{MNorman}, the symmetrized photoemission spectra (EDCs) measured on one Fermi momentum at different temperatures are shown in Fig. 3c.  There is a clear gap opening at low temperatures, as indicated by a dip at the Fermi level in the symmetrized EDCs (Fig. 3c).  With increasing temperature, the dip at E$_F$ is gradually filled up and becomes almost invisible above 40 K. We used two methods to extract the gap size in the superconducting state. The first method is to pick the peak position in the symmetrized EDCs while the second one is to fit the symmetrized EDCs with the phenomenological formula proposed by Norman et al.\cite{MNorman}. The top-middle inset of Fig. 3c shows the symmetrized EDC at 18 K and the fitting result. Both methods give consistent gap size within an experimental uncertainty of 2 meV.   Fig. 3e shows the measured gap size at different temperatures from symmetrized EDCs for both Fermi momenta. The gap size basically follows a BCS form with a T$_c$ around (42$\pm$2) K and a gap size of $\sim$14 meV at zero temperature (Fig. 3e). This is in excellent agreement with the T$_c$$\sim$41 K from direct magnetic measurement (Fig. 3d). We note that FeSe11111 is very sensitive to air and processing temperature. To make sure there is no change on the superconductivity of our sample during the ARPES measurements, we carried out magnetic measurements on the measured sample before and after the ARPES experiment.  Fig. 3d shows the magnetic measurement before ARPES measurement; both the field-cooled (FC) and zero-field-cooled (ZFC) modes give similar magnetic onset T$_c$ $\sim$ 41 K with a sharp transition of $\sim$1.5 K. The sharp transition and its 100\% diamagnetic shielding demonstrate the high quality of the measured single crystal.  After ARPES experiment (not shown in Fig. 3d), the magnetic measurement on the cleaved sample shows an onset T$_c$$\sim$ 40 K with a sharp transition of $\sim$1.5 K. These indicate that there is negligible superconductivity degradation on the sample during the sample handling and ARPES measurements.

Now we come to the momentum dependence of the superconducting gap in the FeSe11111 superconductor.  For this purpose we took high-resolution Fermi surface mapping (energy resolution of 4 meV) of  the electron pocket around M2 at 20 K in the superconducting state (Fig. 4a). The corresponding Fermi momenta are identified, marked by red empty circles, and labelled by numbers from 1 to 16 in Fig. 4a.  Fig. 4b shows symmetrized photoemission spectra (EDCs) at the Fermi momenta around the Fermi surface. The extracted superconducting gap is plotted in Fig. 4c. The superconducting gap  is nearly isotropic with a gap size  at (13$\pm$2) meV; there is no gap node (zero gap) around the Fermi surface.  For comparison, the gap size around the electron-like Fermi surface at M2 point for the  superconducting single-layer FeSe/SrTiO$_3$ film\cite{DFLiu} is also plotted (Fig. 4c). It is also nearly isotropic without gap node although its magnitude ($\sim$15 meV) is slightly larger than that in FeSe11111. In FeSe11111, the bonding between the FeSe layer and the adjacent (Li,Fe)OH layer is rather weak (van der Waals bonding). The separation between the two adjacent FeSe layers is large and their mutual interaction is expected to be rather weak. These make its behaviors close to the single-layer FeSe film that shows strong two-dimensionality. It is thus reasonable to conclude that there is no nodes of the superconducting gap in the FeSe11111 superconductor.

From the above results, we have found that the FeSe11111 superconductor exhibits similar behaviors to the superconducting single-layer FeSe/SrTiO$_3$ film, in terms of the Fermi surface topology, band structure and the superconducting gap symmetry. Amongst all the iron-based superconductors discovered so far, the majority of them show common electronic structure with hole-like Fermi surface sheets around the $\Gamma$ point and electron-like Fermi surface near the M point\cite{Stewart}. The A$_x$Fe$_{2-y}$Se$_2$ superconductor\cite{TQian,DXMouPRL,YZhang,LZhaoPRB,DXReview,Dagotto} and the superconducting single-layer FeSe/SrTiO$_3$ films\cite{DFLiu,SLHe,SYTan,XLReview} have been the only two exceptions that have distinct Fermi surface topology, i.e., no hole-like Fermi surface at the $\Gamma$ point and there are only electron pockets.  The present work has established the FeSe11111 superconductor as a new third exception on the list. It has been proposed that the electron scattering between the hole-like bands near $\Gamma$ and electron-like bands near M is the cause of electron pairing in the iron-based superconductors\cite{Kuroki,Mazin1,FWang1,AVChubukov,VStanev,FWang2,Mazin2}. The discovery of only electron-like Fermi surface in the A$_x$Fe$_{2-y}$Se$_2$ superconductor and the superconducting single-layer FeSe/SrTiO$_3$ films challenges such a Fermi surface nesting scenario. However, some material-specific properties prevent us from making a decisive conclusion. The A$_x$Fe$_{2-y}$Se$_2$ superconductor is well-known for its phase separation problem; the superconducting phase occupies only a small fraction of the entire sample volume and the nature of the superconducting phase remains unclear\cite{YJYan,WBao,ZWang,WLi,DXReview,Dagotto}. The coexistence of the superconducting phase and the insulating 245 phase makes people wonder whether the interface between them is responsible for the observed superconductivity.  On the other hand, the surprising discovery of high temperature superconductivity in the single-layer FeSe/SrTiO$_3$ films indicates that interface may play a significant role in inducing or enhancing superconductivity\cite{QKXue,DFLiu,SLHe,SYTan,WHZhangCPL,JJLee,XLiu,JFHe,RPeng1,RPeng2,JFGe,XLReview}. In this sense, the complications of phase separation and interface effect hinder the direct comparison of the A$_x$Fe$_{2-y}$Se$_2$ superconductors and superconducting single-layer FeSe/SrTiO$_3$ films with most other iron-based superconductors. The newly-discovered FeSe11111 superconductor is advantageous in that it is a single-phase bulk superconductor, free from complications of phase separation and interface effect, thus making it possible to compare directly with other iron-based bulk superconductors. The observation of high temperature superconductivity with a T$_c$ at 41 K in bulk FeSe11111 superconductor that has electron-pocket-only Fermi surface topology makes a strongest and most decisive case that the hole-like Fermi surface around $\Gamma$ is not necessary for the superconductivity in the iron-based superconductors. This rules out Fermi surface nesting scenario as the common pairing mechanism for the iron-based superconductors and points to the significant role of the electron pockets near M in governing high temperature superconductivity in the iron-based superconductors.

The present work also has important implications on the high temperature superconductivity in the single-layer FeSe/SrTiO$_3$ films. First, our results indicate that the electron-pocket-near-M-only Fermi surface topology is not confined to the superconducting single-layer FeSe/SrTiO$_3$ films  only; instead, it can also be present in bulk superconductors. Our experimental results have shown that this form of Fermi surface topology is favorable for achieving high temperature superconductivity in the iron-based superconductors.   Second, it is remarkable that T$_c$ as high as 41 K can be achieved in the bulk FeSe11111 superconductor where there is no strain effect, no interface effect, and no strong electron-phonon coupling effect\cite{JJLee} involved as in the superconducting single-layer FeSe/SrTiO$_3$ films. It is known from the superconducting single-layer FeSe/SrTiO$_3$ films that T$_c$ can be tuned by varying the electron doping level\cite{SLHe}.  Therefore, it is possible that T$_c$=41 K is lower in FeSe11111 than that in the superconducting single-layer FeSe films ($\sim$65 K or higher) because the electron doping level in FeSe11111 is low. This raises a big possibility that T$_c$ of the FeSe11111 superconductor may be further increased if higher electron-doping is made possible and the superconductor-insulator transition\cite{BLei} can be avoided. The high T$_c$ and its possible enhancement with increasing doping in the FeSe11111 superconductor provide key insight on understanding the superconductivity in the single-layer FeSe/SrTiO$_3$ films.   Third, we note that the in-plane lattice constant a or b of the FeSe11111 superconductor (3.78 $\AA$)\cite{XFLuNM,XLDong11111} is comparable to the primary bulk FeSe superconductor (3.76 $\AA$), but significantly smaller than the single-layer FeSe film epitaxially grown on the SrTiO$_3$ substrate (3.905 $\AA$). Considering high T$_c$=41 K and its possible enhancement upon further elecron-doping in the FeSe11111 superconductor, this indicates that the interface strain does not play primary role in determining T$_c$. On the other hand, the distance between the two adjacent FeSe layers seems to have strong correlation with superconductivity in the FeSe-based superconductors. As the distance changes from 5.5 $\AA$ for bulk FeSe, to 9.32 $\AA$ in the FeSe11111 superconductor, and to infinity in the single-layer FeSe/SrTiO$_3$ films, T$_c$ increases from 8.5 K, to 41 K, to 65 K or higher (Fig. 1). In the meantime, the top of the hole-like band at $\Gamma$ changes its position from above the Fermi level for bulk FeSe\cite{JMaletz}, to 50$\sim$70 meV below the Fermi level for FeSe11111, to 80 meV below the Fermi level for the superconducting single-layer FeSe films (Fig. 2). These results again indicate that suppression of the hole-like bands near the $\Gamma$ point is favorable for achieving high temperature superconductivity in the iron-based superconductors.

In summary, by performing high resolution ARPES measurements on the new FeSe-based superconductor, FeSe11111, we have discovered that it exhibits strikingly similar electronic behaviors to the superconducting single-layer FeSe/SrTiO$_3$ film, in terms of the Fermi surface topology, band structure and the superconducting gap symmetry.  This is the first single-phase bulk superconductor with a relatively high T$_c$$\sim$41 K that shows electron-like-Fermi-pockets-only Fermi surface topology that offers a direct comparison with other iron-based bulk superconductors.  Our observations provide significant insights on the superconductivity mechanism in the iron-based bulk superconductors, as well as in the single-layer FeSe/SrTiO$_3$ films.\\

\noindent{\bf\large\textbf{METHODS} }

\noindent{\bf Single crystal preparation.} High quality (Li$_{1-x}$Fe$_x$)OHFe$_{1-y}$Se single crystals were grown by hydrothermal approaches\cite{XLDong11111}. Large crystals of K$_{0.8}$Fe$_{1.6}$Se$_2$ (nominal 245 phase) are specially grown and used as a kind of \emph{matrix} for a hydrothermal ionic exchange reaction. The K ions in K$_{0.8}$Fe$_{1.6}$Se$_2$ are completely released into solution after the hydrothermal reaction process. Single crystal XRD and powder XRD measurements confirm the high quality and high purity of the FeSe11111 single crystals. In particular, no trace of K ions in the FeSe11111 crystals is detected by energy dispersive X-ray spectrometry (EDX) and ICP-AES analyses, confirming a complete release of K ions after the hydrothermal ionic exchange. The sample is sensitive to air and treatment temperature due to the existence of hydroxide.  All the samples for ARPES measurements were prepared in glove box and vacuum case within a short time and then transferred to ARPES measurement chamber to be kept at low temperature during the  ARPES experiments.\\

\noindent{\bf High resolution ARPES measurements.} High resolution angle-resolved photoemission measurements were carried out on our lab system equipped with a Scienta R8000 electron energy analyzer\cite{GDLiu}. We use helium discharge lamp as the light source which can provide photon energies of h$\upsilon$= 21.218 eV (helium I).  The energy resolution was set at 4 meV for both the Fermi surface mapping and band structure measurements (Fig. 1 and Fig. 2) and  for the superconducting gap measurements (Figs. 3 and 4). The angular resolution is $\sim$0.3 degree. The Fermi level is referenced by measuring on a clean polycrystalline gold that is electrically connected to the sample.  The sample was cleaved and measured in vacuum with a base pressure better than 5$\times$10$^{-11}$ Torr.


\vspace{3mm}

\noindent {\bf\large Acknowledgement}\\
We thank Dunghai Lee, Tao Xiang, Guangming Zhang and Zhongyi Lu for discussions.  X.J.Z. thanks financial support from the NSFC (11190022, 11334010 and 11374335), the MOST of China (973 program numbers: 2011CB921703, 2015CB921000 and 2011CBA00110) and the Strategic Priority Research Program (B) of CAS with Grant No. XDB07020300.  X.D., F. Z. and Z.X.Z thank financial support from the NSFC(11274358), the MOST of China (973 program number 2013CB921701) and the Strategic Priority Research Program (B) of CAS with Grant No. XDB07020100.
\vspace{3mm}

\noindent {\bf\large Author Contributions}\\
 X.J.Z., L.Z., X.L.D. and Z.X.Z. proposed and designed the research. D.N.Y., H.X.Z, Y.L.H., X.L.D., F.Z. and Z.X.Z.contributed to single crystal preparation. L.Z., A.J.L., Y.H., D.F.L., J.W.H., S.L.H., B.S., Y.X., X.L., L.Y., G.D.L., X.L.D., Z.Y.X., C.T.C.and X.J.Z. contributed to the development and maintenance of Laser-ARPES system. L.Z., A.J.L., Y.H. carried out the experiment with D.F.L., J.W.H., S.L.H., B.S., Y.X., and X.L.. L.Z., A.J.L., Y.H. and X.J.Z. analyzed the data. X.J.Z. and L.Z. wrote the paper. All authors discussed the results and commented on the manuscript.

\vspace{3mm}

\noindent {\bf\large Additional information}\\
\noindent{\bf Competing financial interests:} The authors declare no competing financial interests.

\newpage

\begin{figure}[tbp]
\begin{center}
\includegraphics[width=0.90\columnwidth,angle=0]{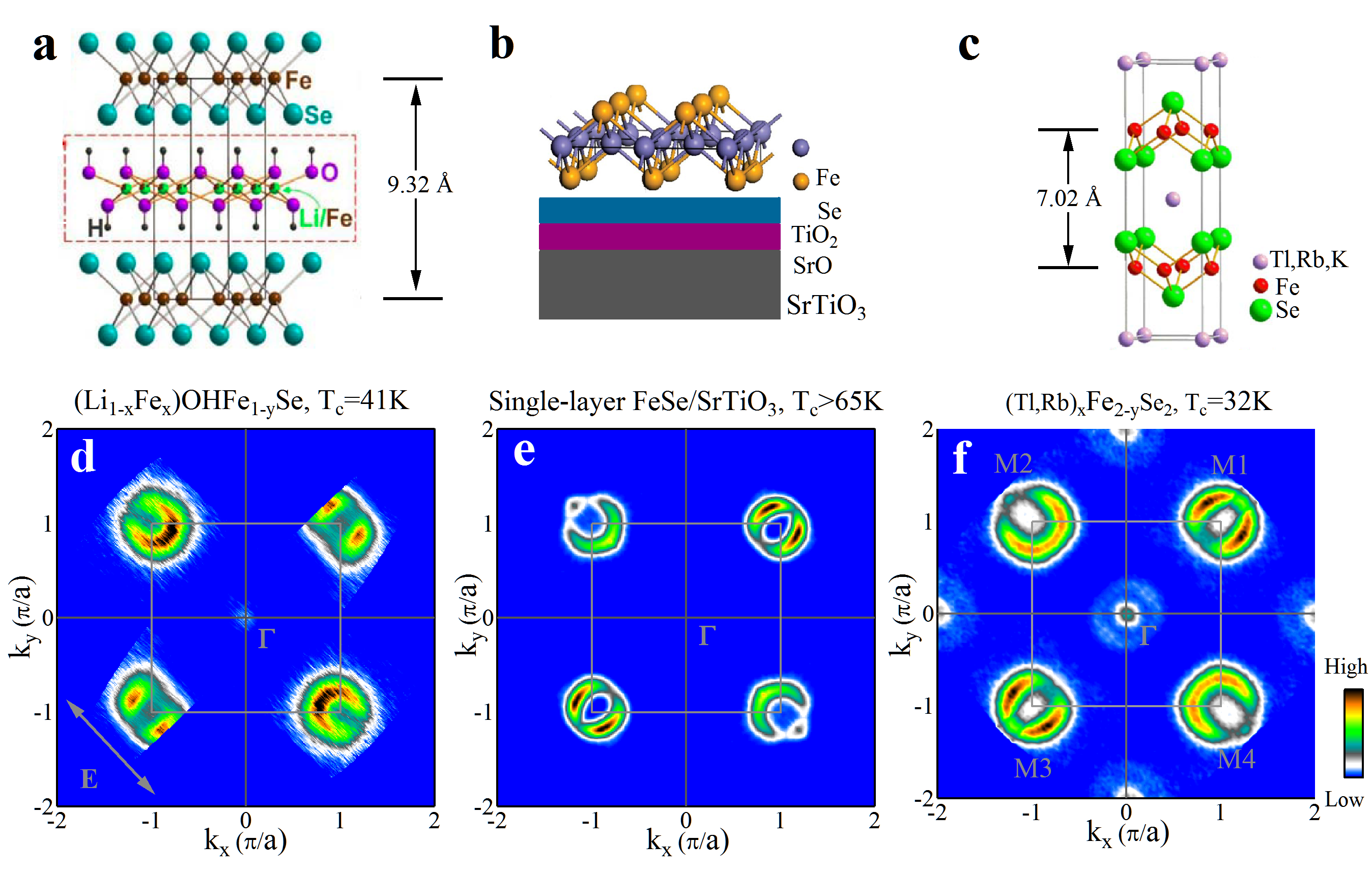}
\end{center}
\caption{{\bf Fermi surface of (Li$_{0.84}$Fe$_{0.16}$)OHFe$_{0.98}$Se superconductor and its comparison with that in the superconducting single-layer FeSe/SrTiO$_3$ film and (Tl,Rb)$_x$Fe$_{2-y}$Se$_2$ superconductor.} (a). The crystal structure of (Li$_{1-x}$Fe$_x$)OHFe$_{1-y}$Se. The {\it c} lattice constant is 9.318 ${\AA}$ for optimally-doped sample with a T$_c$ at 42 K\cite{XFLuNM,XLDong11111}.   (b). Schematic structure of single-layer FeSe film deposited on a SrTiO$_3$ substrate\cite{QKXue}. The thickness of the FeSe layer is $\sim$5.5 ${\AA}$. (c). Schematic structure of (Tl,Rb)$_x$Fe$_{2-y}$Se$_2$\cite{JGGuo}. The distance of the adjacent FeSe layers is $\sim$7.02 ${\AA}$. (d). Fermi surface mapping of (Li$_{0.84}$Fe$_{0.16}$)OHFe$_{0.98}$Se superconductor measured at 20 K. The spectral weight distribution was obtained by integrating the measured photoemission spectra (EDCs) within [-10,+10] meV energy window with respect to the Fermi level as a function of k$_x$ and k$_y$.  (e)-(f). Fermi surface of a superconducting single-layer FeSe/SrTiO$_3$ film\cite{DFLiu} and (Tl,Rb)$_x$Fe$_{2-y}$Se$_2$ \cite{DXMouPRL} . For convenience, the four equivalent M points are labeled as M1($\pi$,$\pi$), M2(-$\pi$,$\pi$), M3(-$\pi$,-$\pi$) and M4($\pi$,-$\pi$). The light-grey arrow in (d) marks the main electric field direction on the sample surface from the light source. In (d) to (f), the data around M1 and M4 are obtained from the data around M3 and M2, respectively, by applying the inversion symmetry.
}
\end{figure}

\begin{figure*}[tbp]
\begin{center}
\includegraphics[width=0.95\columnwidth,angle=0]{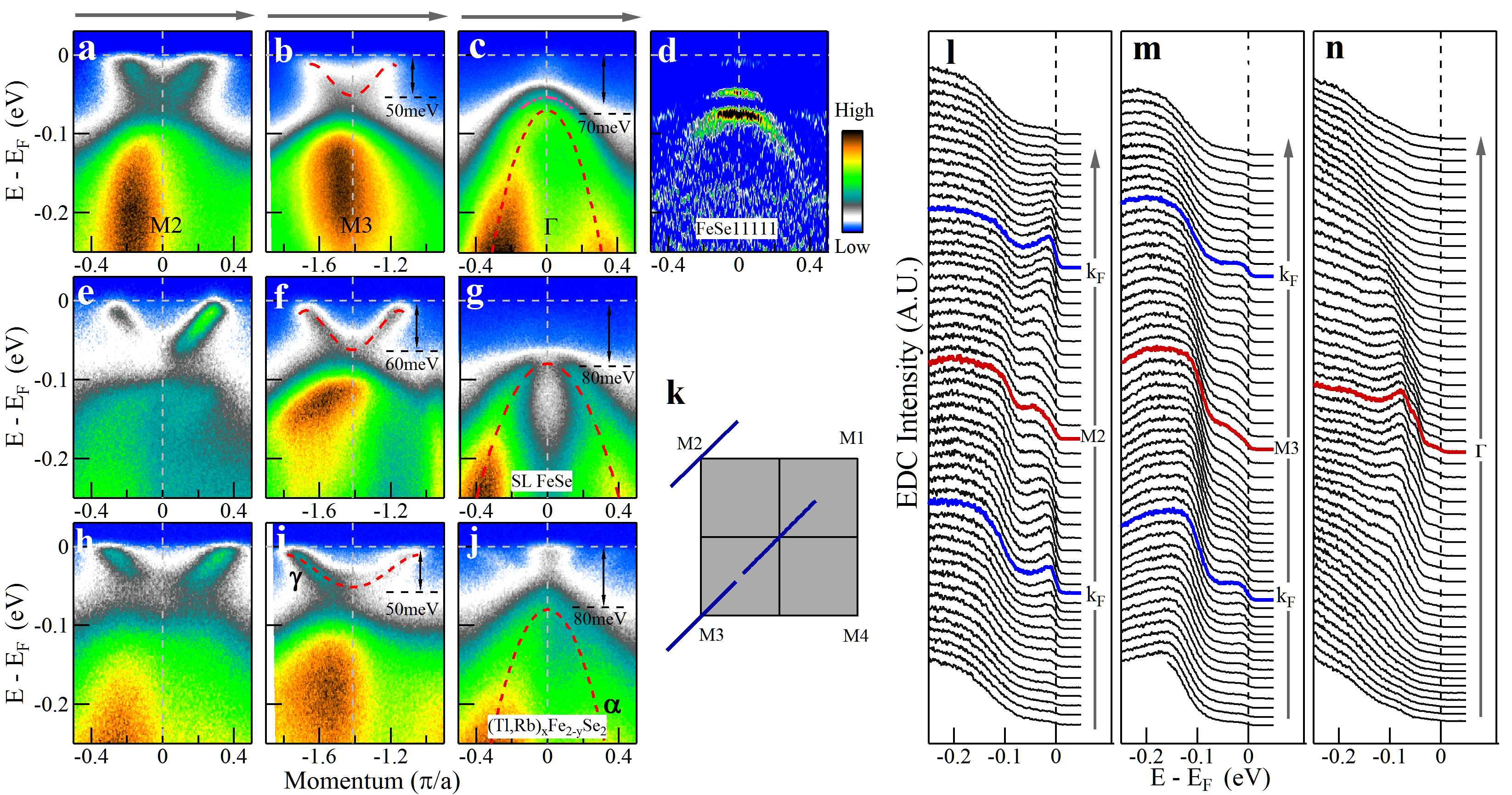}
\end{center}
\caption{{\bf Band structure and photoemission spectra of (Li$_{0.84}$Fe$_{0.16}$)OHFe$_{0.98}$Se superconductor measured along high-symmetry momentum cuts.} (a) to (c) show photoemission images of (Li$_{0.84}$Fe$_{0.16}$)OHFe$_{0.98}$Se measured at a temperature of 20 K along three high-symmetry cuts, $\Gamma$ cut, M2 cut and M3 cut, respectively. (d) shows the second-derivative image of (c) with respect to the energy in order to reveal the band structure better. For comparison, (e) to (g) and (h) to (j) show photoemission images of a superconducting single-layer FeSe/SrTiO$_3$ film and (Tl,Rb)$_x$Fe$_{2-y}$Se$_2$ measured along the same $\Gamma$ cut, M2 cut and M3 cut, respectively.  The location of the three momentum cuts are shown in (k).  The red dashed lines in (a-j) images are the guide to the eyes for the observed band structures.  (l), (m) and (n) show photoemission spectra (energy distribution curves, EDCs) of  (Li$_{0.84}$Fe$_{0.16}$)OHFe$_{0.98}$Se for the $\Gamma$ cut, M2 cut and M3 cut, corresponding to (a-c) images, respectively. The photoemission spectra at $\Gamma$ point are colored as red while the spectra at the Fermi momenta (k$_F$s) are colored as blue. }
\end{figure*}

\begin{figure}[tbp]
\begin{center}
\includegraphics[width=0.76\columnwidth,angle=0]{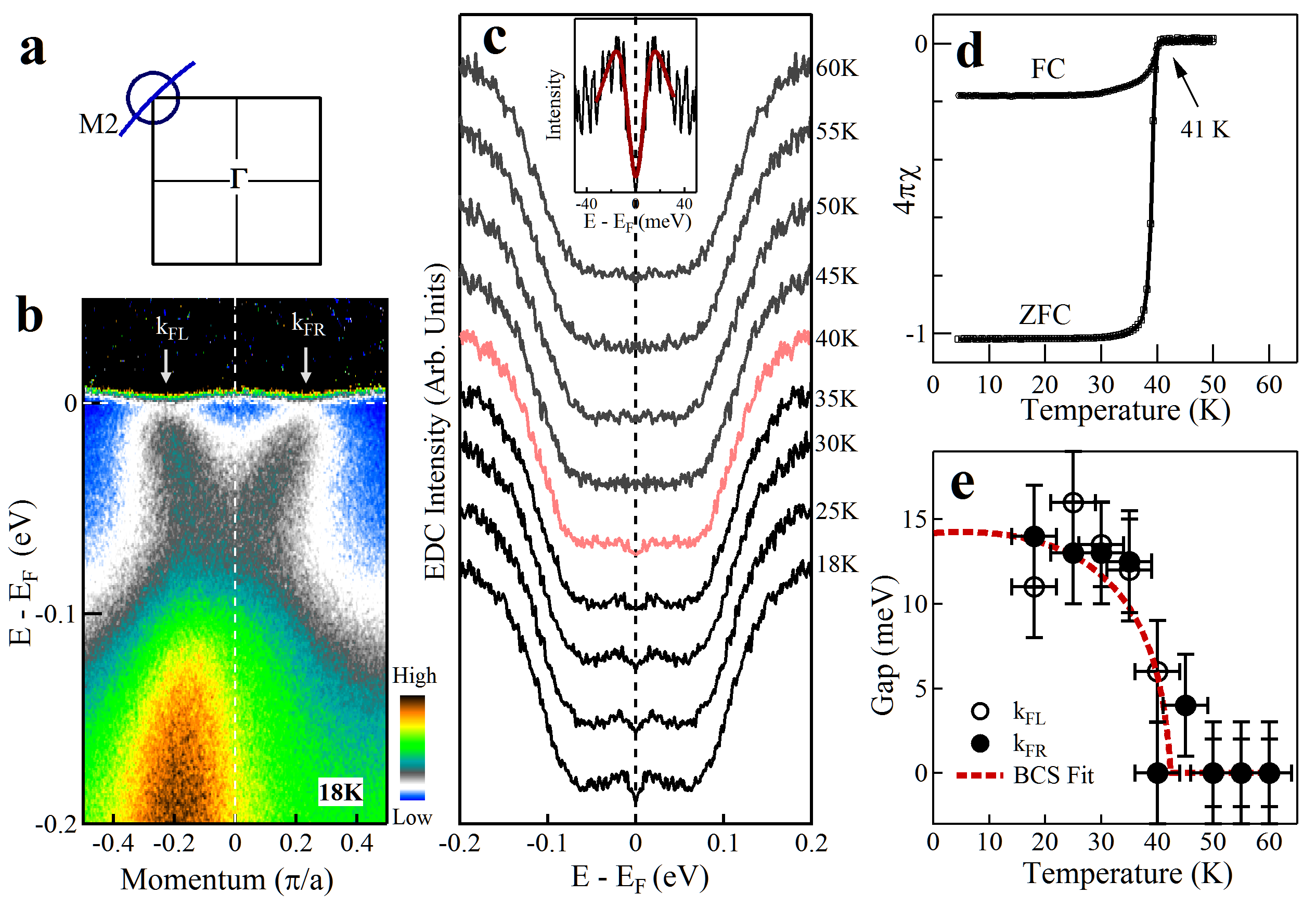}
\end{center}
\caption{{\bf Temperature dependence of the superconducting gap of the (Li$_{0.84}$Fe$_{0.16}$)OHFe$_{0.98}$Se superconductor.} (a) shows schematically the electron-like Fermi surface near M2 point and the location of the momentum cut (blue line) across the Fermi surface. (b). Photoemission image measured at 18 K along the momentum cut near the M2 point (blue line in (a)).  The data is divided by the corresponding Fermi distribution function in order to highlight the opening of an energy gap around the Fermi level. Two Fermi momenta are marked by arrows as k$_{FL}$ and k$_{FR}$. (c). Symmetrized photoemission spectra (EDCs) at the Fermi momentum k$_{FL}$ measured at different temperatures. The middle-top inset shows the expanded region of the 18 K symmetrized EDC around the Fermi level and the fitting result (red line) using the formula from Ref.\cite{MNorman} to get the gap size.  (d). Magnetic measurement of the superconducting transition temperature (T$_c$) for the (Li$_{0.84}$Fe$_{0.16}$)OHFe$_{0.98}$Se sample we measured. Both field-cooled (FC) and zero-filed-cooled (ZFC) mode measurements  show an onset T$_c$ at $\sim$41 K with a sharp transition width of $\sim$1.5 K.  (e).  Temperature dependence of the measured superconducting gap. The gap size is obtained by picking up the peak position from the symmetrized EDCs (c) or by fitting the near-E$_F$ symmetrized EDCs using the formula from Ref.\cite{MNorman}; both approaches give similar results within experimental uncertainty.  The error bars are defined as standard deviation.  The gap size from EDCs of both the left Fermi momentum (k$_{FL}$) and the right Fermi momentum (k$_{FR}$) are plotted. The red dashed line is a BCS gap form with a gap size  of $\sim$14 meV at zero temperature.
}
\end{figure}

\begin{figure}[tbp]
\begin{center}
\includegraphics[width=0.85\columnwidth,angle=0]{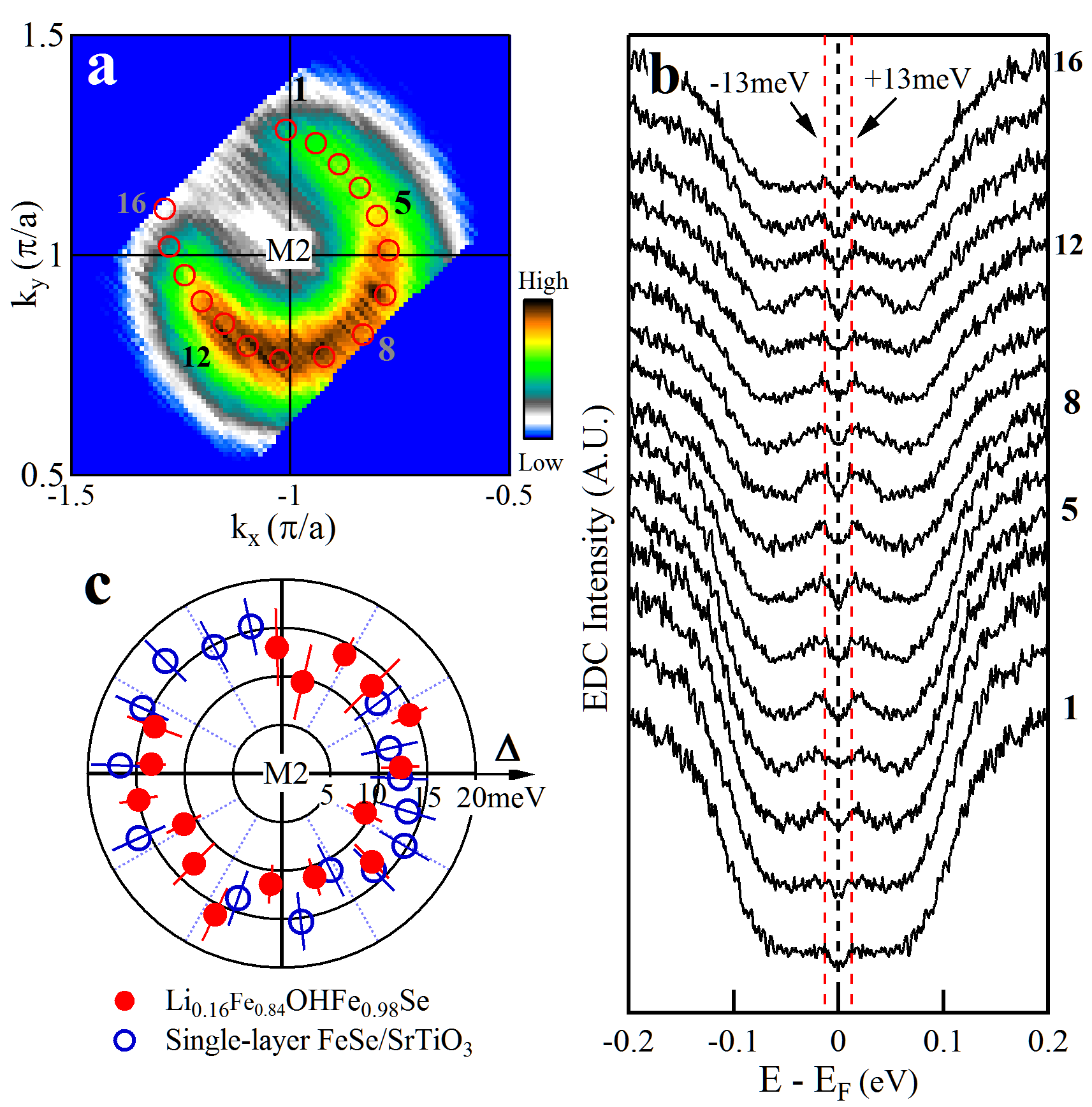}
\end{center}
\caption{{\bf Momentum dependence of the superconducting gap of (Li$_{0.84}$Fe$_{0.16}$)OHFe$_{0.98}$Se superconductor.}  (a). Fermi surface mapping near M2 point. The corresponding Fermi crossings are marked by red empty circles with numbers from 1 to 16 (for clarity, the intermittent points are not labeled). (b). Symmetrized EDCs measured at 20 K from Fermi crossing 1 to 16 along the measured electron-like Fermi surface around M2. The red dashed lines represent the energy positions of $\pm$ 13 meV with respect to the Fermi level. (c). Momentum dependence of the superconducting gap (red solid circles) along the Fermi surface around M2 for (Li$_{0.84}$Fe$_{0.16}$)OHFe$_{0.98}$Se superconductor.  For comparison, the gap along the electron-like Fermi surface around M2 of the superconducting single-layer FeSe film is also plotted (blue empty circles)\cite{DFLiu}. The error bars are defined as standard deviation.
}
\end{figure}

\end{document}